\documentclass[preprint2]{aastex}
\usepackage{graphicx,multirow}
\DeclareGraphicsExtensions{.eps}

\begin{document}
\title{Red Clump Stars in the Sagittarius Tidal Streams}
\author{Kenneth Carrell\altaffilmark{1,2,3}, 
Ronald Wilhelm\altaffilmark{4} and
YuQin Chen\altaffilmark{1}}
\email{carrell@nao.cas.cn}

\altaffiltext{1}{Key Laboratory of Optical Astronomy, 
National Astronomical Observatories, Chinese 
Academy of Sciences, Beijing 100012, China}
\altaffiltext{2}{Physics Department, Texas Tech University, Lubbock, 
Texas 79409}
\altaffiltext{3}{Visiting Astronomer, Kitt Peak National 
Observatory, which is operated by the Association of 
Universities for Research in Astronomy, Inc. (AURA) under 
cooperative agreement with the National Science Foundation}
\altaffiltext{4}{Physics and Astronomy Department, University of 
Kentucky, Lexington, KY 40506}

\begin{abstract}
We have probed a section ($l$~$\sim$~150,~$b$~$\sim$~-60) of 
the trailing tidal arm of the Sagittarius dwarf spheroidal 
galaxy by identifying a sample of Red Clump stream stars.  
Red Clump stars are not generally found in the halo field, but 
are found in significant numbers in both the Sagittarius galaxy 
and its tidal streams, making them excellent probes of stream 
characteristics.  Our target sample was selected using 
photometric data from the Sloan Digital Sky Survey, Data Release 
6, which was constrained in color to match the Sagittarius Red 
Clump stars.  Spectroscopic observations of the target stars 
were conducted at Kitt Peak National Observatory using the 
WIYN\footnote{The WIYN Observatory is a joint facility of the 
University of Wisconsin-Madison, Indiana University, Yale 
University, and the National Optical Astronomy Observatories.}
telescope.  The resulting spectroscopic sample is magnitude 
limited and contains both main sequence disk stars and evolved 
Red Clump stars.  We have developed a method to systematically 
separate these two stellar classes using kinematic information 
and a Bayesian approach for surface gravity determination.  The 
resulting Red Clump sample allows us to determine an absolute 
stellar density of $\rho$~=~2.7~$\pm$~0.5~RC~stars~kpc$^{-3}$ at 
this location in the stream.  Future measurements of stellar 
densities for a variety of populations and at various locations 
along the streams will lead to a much improved understanding of 
the original nature of the Sagittarius galaxy and the physical 
processes controlling its disruption and subsequent stream 
generation.
\end{abstract}

\keywords{galaxy: halo --- galaxies: interactions ---
stars: horizontal branch}

\section{Introduction}\label{sec:intro}
Since its discovery \citep{iba94}, the Sagittarius dwarf spheroidal 
galaxy (Sgr) has served as an important probe for the understanding of 
the properties of our Galaxy.  Unlike other previously identified 
Milky Way satellite galaxies, the Sgr is a relatively nearby galaxy 
with clear evidence of being tidally disrupted.  Less than two 
years after its discovery the Sgr was found to have 'tidal tails' 
\citep{mat96} - stars which have been stripped from the main 
body of the dwarf galaxy through its interaction with the Milky Way 
Galaxy.  This led to the finding that the Sgr actually had an 
extended and prominent stream of stars that was ultimately traced 
over the entire sky \citep{maj03}.

The core of the Sgr is located below the bulge of our Galaxy 
as seen from the location of the Sun.  Its orbit has been found 
to be near polar and both leading and trailing streams of 
tidally stripped stars accompany the core.  These streams trace 
out the full orbit of the Sgr and overlap each other at several 
locations (see, for example, the models of 
\citet{joh05,law05,fel06,law10}).  It is currently not known how 
many orbits are represented in the streams, how long ago the stars 
were liberated from their host, the Sgr initial mass function (IMF), or 
the initial internal dynamics of the Sgr.  All of these are important
parameters that are needed to better constrain the Sgr orbital models.
As these models become more realistic much stronger contraints will be
placed on the shape of the dark matter halo and the disruption processes
that govern dwarf satellite accretion (see 
\citet{gom99,iba01,hel01,hel04a,hel04b,joh05,fel06} 
and references therein).

In the last decade, much progress has been made in characterizing 
properties of the streams.  The 2MASS and SDSS survey programs have 
provided a wealth of information on stream location and kinematics 
using various stellar populations.  Specifically, giants 
\citep{bel03,maj03} and turnoff stars \citep{new02} were among the 
first to provide a wider view of the full system, while more 
recent work has been concentrated on the more evolved stars of 
the horizontal branch (HB), both on the blue 
\citep{yan09,ruh11} and red \citep{cor10,shi12} ends.  The HB provides 
some benefits not available in other stellar populations.  The blue 
HB extends blueward of the thick-disk and halo main-sequence turnoff 
making sample selection much easier compared to stars redward of the 
turnoff.  Furthermore, placement on the HB is dependent on the 
age and metal abundance of the star - those on the blue end are older 
(and, in general, more metal poor) with decreasing age (and, in 
general, increasing abundance) as a function of redder color.  The 
HB terminates on the red end with a red clump of stars where the 
mechanism that drives the HB placement is no longer as effective in 
changing the color of the He-core burning star.  This age/metallicty 
relation between the older, blue, metal-poor HB and the younger, 
metal-rich, RC can be exploited to probe changes in the contribution 
of stellar populations in the streams and Sgr proper.

In particular, a photometric study by \citet{bel06} found a gradient 
in the ratio of old to young stars along the arms of the Sgr tidal 
streams.  This study compared the ratio of BHB stars to RC stars 
in the core of the Sgr and in an area far from 
the core along one of the tidal streams.  The authors found the 
BHB/RC ratio to be five times larger in the stream than in the 
core of the dwarf galaxy.  This led \citet{bel06} 
to conclude that a steep, radial, age/metallicity gradient must 
have been present in the progenitor Sgr prior to the tidal stripping events 
that produced the stream.  This 
result was based solely on star counts from photometric data 
where contamination from various sources (galaxies in the 
BHB colors and foreground disk stars in the RC colors) were 
difficult to eliminate.  

This current paper is part of a program to observe   
RC stars spectroscopically and to confirm their stream membership
through kinematics and surface gravity determinations.    
This method greatly reduces the effects of background 
contamination.  A clean sample of RC stream members is used to 
determine the stellar density of stars at one point 
along the Sgr stream.  This program, coupled 
with stellar densities from other populations, ultimately holds 
the potential for the reconstruction of the Sgr IMF, initial 
stellar distribution, and clues to the internal dynamics of the progenitor
Sgr.

In section \ref{sec:data} of this paper we will discuss the sample 
selection criteria and the observations and reductions of our 
spectroscopic data.  Section \ref{sec:analysis} describes the 
analysis procedure used to provide a clean RC sample.  In section 
\ref{sec:results} we present the results found using this stellar 
sample and follow it up in section \ref{sec:discussion} with a 
discussion of these results.

\section{Data}\label{sec:data}
\subsection{The Sample}\label{ssec:samp}
Our candidate stars were selected using 
the Sloan Digital Sky Survey (SDSS) online database\footnote{
\url{http://www.sdss.org}}.  Using 
previous measurements \citep{maj03,maj04} and simulated 
models \citep{law05,fel06,law10} of the Sgr tidal tails, a dynamically 
cold and kinematically distinct region of the stream was chosen 
for study.  The kinematic signature will aid in eliminating 
contaminants in our color selection (disk dwarfs and halo giants).
We chose to observe a region of the trailing arm at
($l$,$b$) of approximately (150$^\circ$,-60$^\circ$).  This region 
had the advantages of being previously studied by other investigators
and a known stream distance (less than 40~kpc) which fell within the 
limiting magnitude ($g$~$\sim$~19) for our current project. In addition, the 
models of \citet{law05} predict that the more dispersed leading arm 
overlaps in distance at this location.  This afforded us the possibilty
of detecting the leading arm within the same dataset.

\begin{figure}[tb]
\plotone{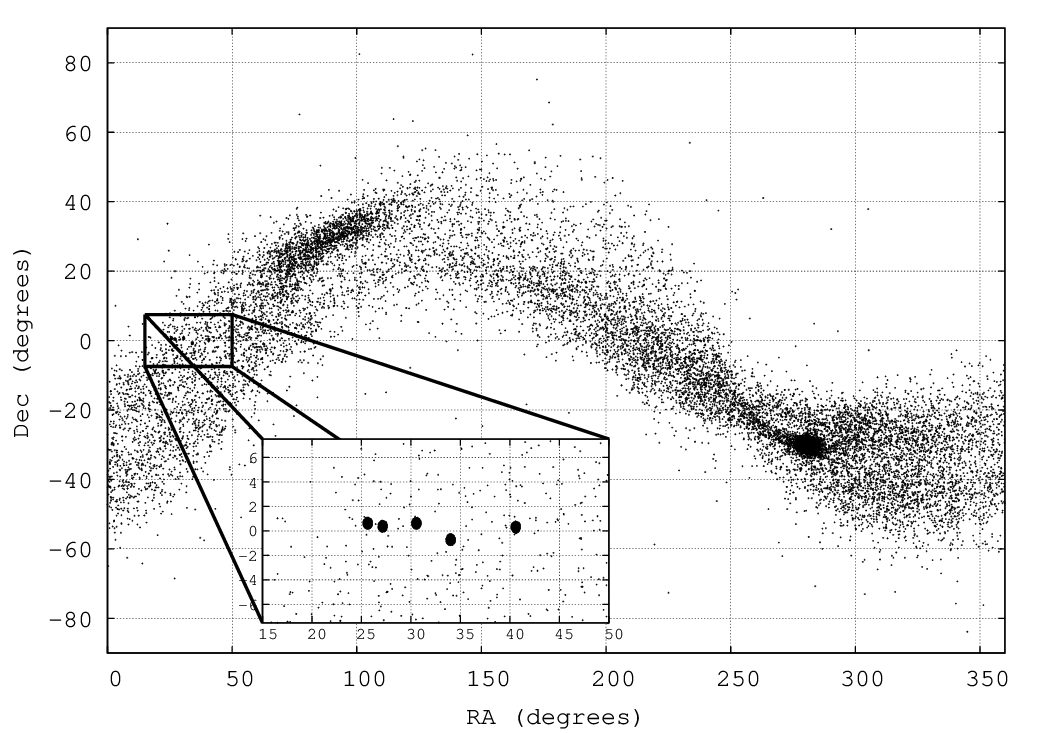}
\caption{Equatorial coordinates for the triaxial halo model of 
\citet{law10} and the fields observed in this research.  The filled 
circles in the inset correspond to observed fields.  The core of 
the Sgr dwarf galaxy is the clump of stars located at 
($RA$,$Dec$)$\sim$(285$^\circ$,-30$^\circ$).}
\label{fig:fieldpos}
\end{figure}
The expected stream distances at this location are $\sim$~30~kpc 
and the streams should be dynamically cold with predicted velocities 
for the leading (trailing) arm of roughly +(-)150~km~s$^{-1}$.  These 
velocities are uniquely different from the larger number of 
contaminating disk stars which have velocities of nearly 
0~km~s$^{-1}$ in this direction and the halo which has an average 
velocity slighly less than -100~km~s$^{-1}$ in this direction and 
has a large velocity dispersion.  This insured that stream 
identification could be made, in part, from the kinematics of the 
RC stars.  Table \ref{tab:fields} shows relevant values for our 
selected fields and Figure \ref{fig:fieldpos} shows our field 
positions in right ascension and declination overlaid on the 
triaxial halo model of \citet{law10}.
\begin{table}[tb]
\begin{center}
\caption{Field Locations\label{tab:fields}}
\begin{tabular}{ c c c c c }
\tableline
\multirow{2}{*}{{\bf Field}} & {\bf RA} & {\bf Dec} & {\bf $l$} & {\bf $b$} \\
 & (degrees) & (degrees) & (degrees) & (degrees) \\
\tableline
f02 & 27.15 & +0.383 & 151.8 & -59.25 \\
f05 & 25.63 & +0.633 & 148.9 & -59.64 \\
f09 & 30.55 & +0.633 & 157.3 & -57.44 \\
f11 & 34.02 & -0.717 & 164.0 & -56.68 \\
f18 & 40.60 & +0.317 & 171.7 & -51.74 \\
\tableline
\end{tabular}
\end{center}
\end{table}

To choose our candidate stars, color and magnitude cuts 
were determined that would encompass the Sgr RC stars while minimizing 
contamination from stars in the disk of our Galaxy.  To establish
the color boundaries for the RC in the Sgr galaxy we used the 
extinction corrected $V-I$ photometry data of \citet{lay00}.  
We then transformed these color boundaries to the SDSS photometric 
system using photometric transformation equations by Robert Lupton
on the SDSS website\footnote{
\url{http://www.sdss.org/dr6/algorithms/sdssUBVRITransform.html}}.  
The magnitude range was influenced by the expected distances of 
stream stars (30~kpc corresponds to a $g$ magnitude of slightly 
fainter than 18 assuming an absolute magnitude of $g_{abs}\sim$1.0), 
telescope aperture, and the reliable range of the 
SDSS photometry.  Our final adopted selection criteria were 
0.5~$<$~$(g-r)_\circ$~$<$~0.65 and 15.5~$<$~$g_\circ$~$<$~18.7. 
\begin{figure}[tb]
\plotone{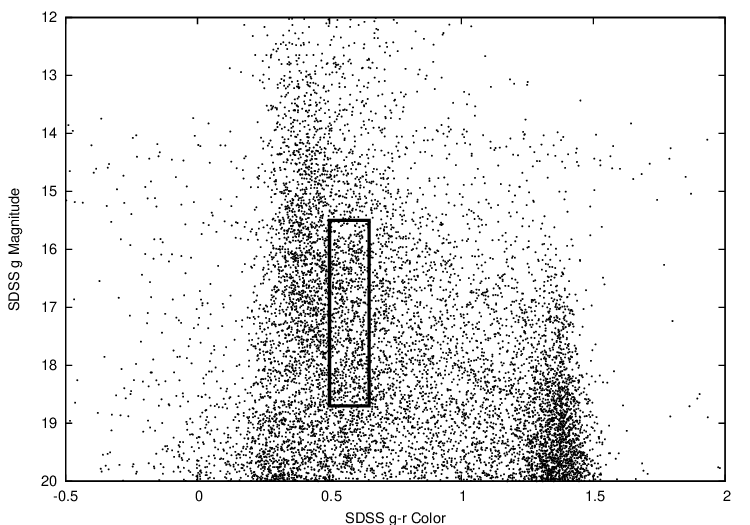}
\caption{A representative color-magnitude diagram of stars in 
our selected area of the sky.  The box represents the selection 
criteria used to select the RC program stars 
of 0.5~$<$~$(g-r)_\circ$~$<$~0.65 and 15.5~$<$~$g_\circ$~$<$~18.7.}
\label{fig:CMD}
\end{figure}
Figure \ref{fig:CMD} is a representative $g_\circ$ versus $(g-r)_\circ$ 
CMD for candidate stars in one of our fields with the selection 
box overlaid.  The range in magnitude corresponds to a distance 
range of roughly 8 to 35~kpc (assuming an absolute magnitude of 
$g_{abs}\sim$1.0).  This magnitude range insured the inclusion
of all RC stars in the primary trailing arm and also allowed for 
the inclusion of other, nearer portions of the stream that are 
suggested from the stream models.

While the chosen color and magnitude range encompasses the candidate 
RC stars, it also allows for the inclusion of a significantly large number of 
contaminating main-sequence stars from the Galactic disk.  In addition to the
disk stars there is also a relatively small number of K giants that are 
expected from both the halo field and the Sgr stream.  To minimize the 
contamination from both the disk and halo field we used a combination 
of kinematics and computed surface gravities.  The kinematic signature 
from the dynamically cold stream is easily distinguished from both the 
disk and halo populations.  Furthermore, the surface gravity for the 
RC stars is significantly lower than the main-sequence stars in the disk.
By combining both the kinematic information and the stellar surface gravity
determinations we are able to get a clean sample of stream giant stars.

\subsection{The Observations}\label{ssec:obs}
The spectroscopic observations were taken at the WIYN telescope at KPNO 
using the Hydra instrument.  Observations were carried out in August 2007.  
Hydra has 83 active blue fibers that have a diameter of 3.1~arc-seconds 
on the sky and are positioned within the 1$^\circ$ diameter field of 
view.  The Simmons Camera with no filter and a 600 lines per mm grating 
with a central wavelength of 5500~\AA\ was used.  This gave a resolution 
of approximately 4.6~\AA\ and a dispersion of just over 1~\AA\ per pixel.  
Our detector setup provided us with a wavelength coverage of roughly 
4000-5500~\AA.  The density of candidates and the positioning 
constraints of the instrument allowed on average 42 objects to be 
simultaneously observed in each field.

For each program field, four, 30 minute integrations were obtained, 
cleaned of cosmic-ray events and co-added, for a total integration time
of two hours per field. This exposure time resulted in data with a typical 
SNR of $\sim$~45 at a $g_\circ$ magnitude of 16.5 and $\sim$~20 at a $g_\circ$ 
magnitude of 18.

The reduction of all images was performed using the IRAF 
software package\footnote{\url{http://iraf.noao.edu}}.  Bias level 
subtraction, flat-field division using dome flats, sky subtraction, 
spectral extraction, and wavelength calibrations from a CuAr 
lamp were all performed using the {\it dohydra} routine in IRAF.

\section{Analysis}\label{sec:analysis}
In an effort to extract meaningful information from the spectral 
data obtained as described in section \ref{sec:data}, an analysis 
software framework was developed by the authors.  The ability to 
separate the RC stream stars from main sequence disk stars hinged 
on the ability to determine the kinematics of our sample as well 
as reliable stellar parameters.  Of particular interest is the 
ability to quantify surface gravity since this parameter is a
key discriminate for the RC sample. 

\subsection{The Procedure}\label{ssec:proc}
Our procedure uses a grid of synthetic spectra to compare to
our observed data.
The synthetic stellar spectra were generated using the 
SPECTRUM program\footnote{Developed by Richard Gray, 
\url{http://www.phys.appstate.edu/spectrum/spectrum.html}} 
with ATLAS9 model atmospheres\footnote{From Robert Kurucz, 
\url{http://kurucz.harvard.edu}} as input.  These grids 
were generated to encompass the entire parameter space 
likely to be observed in the program stars.  The grid size 
for the effective temperature was 4000~to~6000~K with a step 
size of 250~K.  For surface gravity a step size of 0.5~dex was 
used over a range of 0.5~to~5.0~dex.  The metallicity range was 
-2.5~to~+0.5~dex in increments of 0.5~dex.

A primary concern we had with the synthetic data (as well as 
the scientific data) was how the normalization would be 
carried out.  While SPECTRUM 
allows for the synthetic spectra to be output in a normalized form, 
this true normalization can be quite different from the normalized, 
pseudo-continuum
found for the science data by fitting the non-flux calibrated spectra. 
Direct comparison between the two normalizatons leaves open the 
possibility that systematic 
errors could be introduced in the form of incorrect line strengths 
because of incorrect continuum placement.  To avoid this possible 
problem, the synthetic spectra were output with absolute flux 
values and both these and the observed spectra were normalized by 
fitting the continuum with a third order polynomial.

\begin{figure}[bt]
\plotone{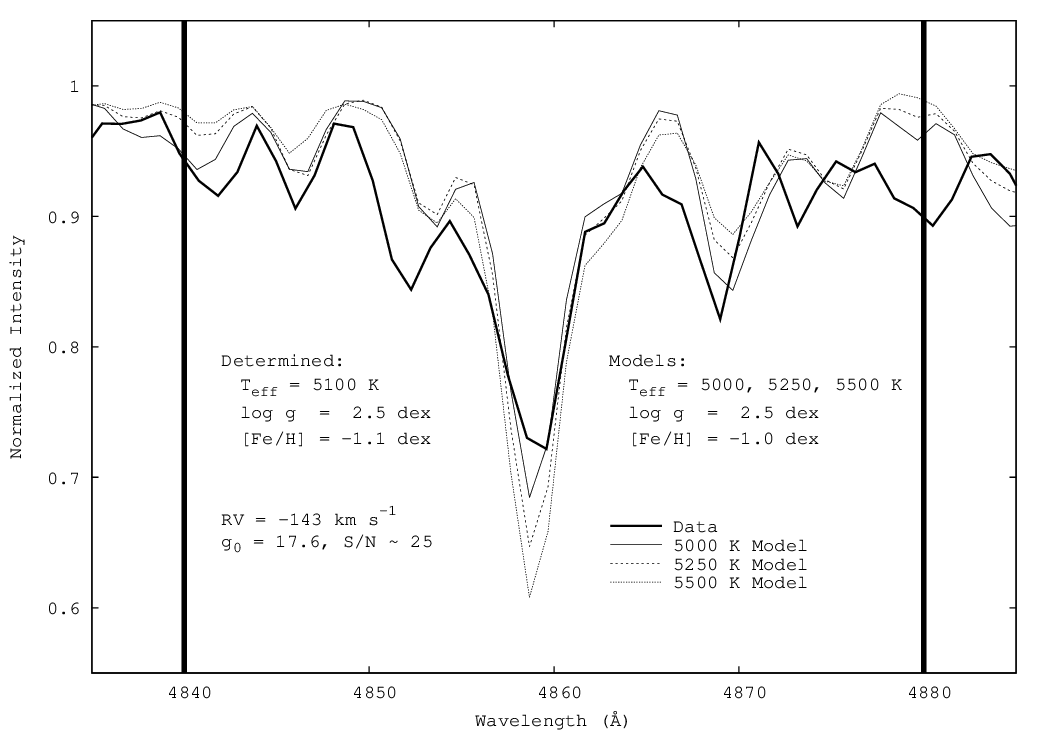}
\caption{Normalized example spectrum of an observed star showing the 
wavelength range used to determine the effective temperature (marked 
with vertical lines).  Spectroscopic data is shown with a thick 
solid line.  The synthetic data has been shifted to match the 
determined radial velocity of the star.  The major feature in this 
range is the H$_{\beta}$ line.}
\label{fig:teff}
\end{figure}

\begin{figure}[bt]
\plotone{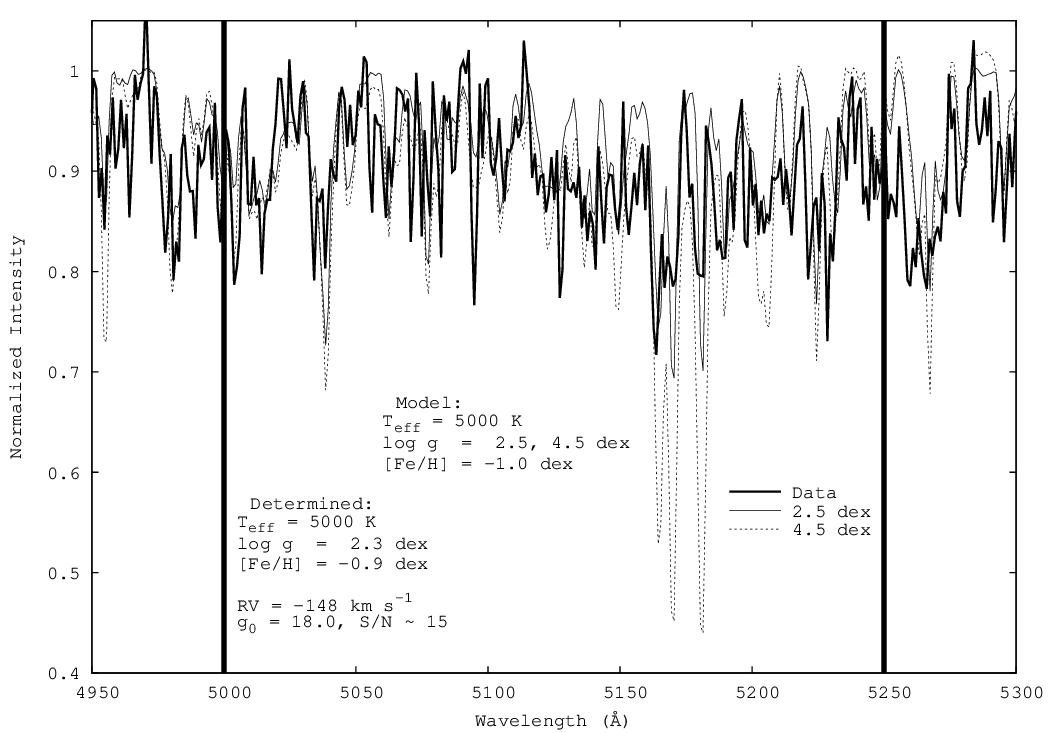}
\caption{Normalized example spectrum of an observed star showing the 
wavelength range used to determine the surface gravity (marked with 
vertical lines).  Spectroscopic data is shown with a thick solid 
line.  The overlaid synthetic spectra have been shifted to match the 
determined radial velocity of the star.  The magnesium b lines are 
the dominant feature in this wavelength range.}
\label{fig:logg}
\end{figure}

\begin{figure}[bt]
\plotone{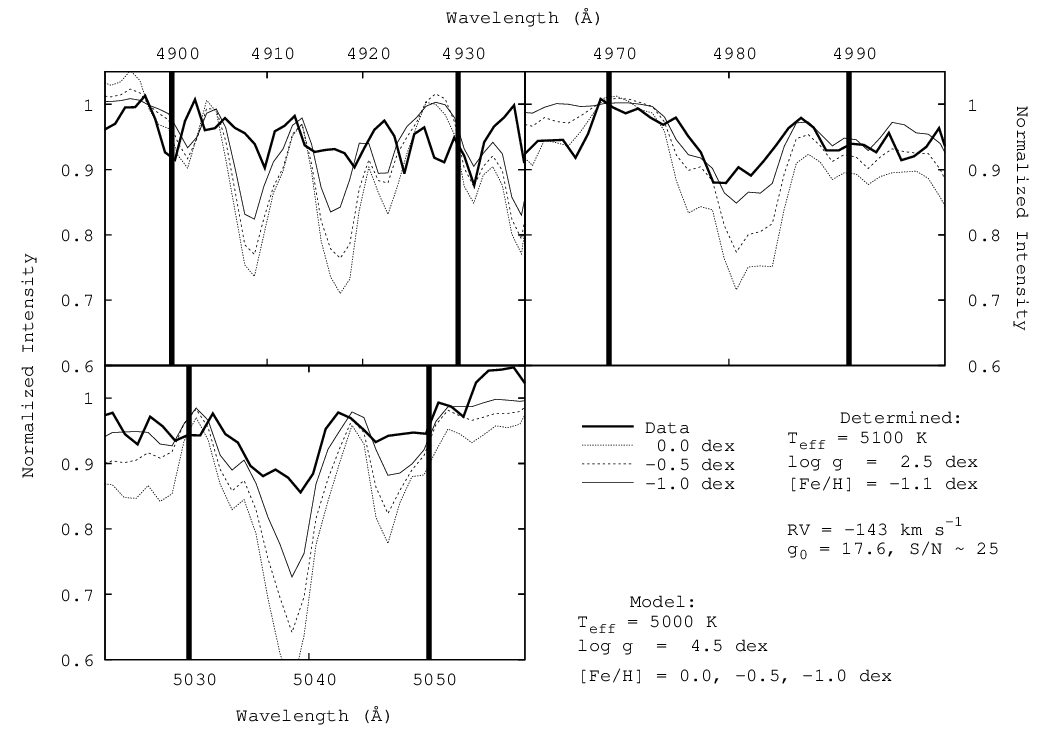}
\caption{Normalized example spectrum of an observed star showing 
three of the wavelength ranges used to determine the metallicity 
(marked with vertical lines).  Spectroscopic data is shown with a 
thick solid line.  The overlaid synthetic spectra have been shifted 
to match the determined radial velocity of the star.  These 
wavelength ranges include primarily Fe lines.}
\label{fig:abun}
\end{figure}

A table of the wavelength ranges used to determine each parameter 
can be found in Table \ref{tab:ranges}.
\begin{table}[tb]
\begin{center}
\caption{Wavelength Regions for Parameter Determinations\label{tab:ranges}}
\begin{tabular}{ c c }
\tableline
{\bf Wavelength Range} & \multirow{2}{*}{{\bf Parameter}}\\
(\AA) & \\
\tableline
4840 - 4880 & T$_{eff}$ \\
\tableline
5000 - 5250 & $\log g$ \\
\tableline
4360 - 4400 & [Fe/H] \\
4630 - 4680 & [Fe/H] \\
4900 - 4930 & [Fe/H] \\
4970 - 4990 & [Fe/H] \\
5030 - 5050 & [Fe/H] \\
5220 - 5290 & [Fe/H] \\
5435 - 5465 & [Fe/H] \\
\tableline
\end{tabular}
\end{center}
\end{table}
We are using strictly the H$_{\beta}$ line as a temperature indicator and 
the wavelength region that includes the magnesium b lines and magnesium 
hydride molecule as a surface gravity indicator.  Other areas 
were investigated for both parameters but nothing was found to 
improve the results from these single ranges.  Our abundance 
information is spread over the entire observed wavelength range but 
excludes the previously mentioned areas used for other parameters.  
In Figures \ref{fig:teff}, \ref{fig:logg}, and \ref{fig:abun} 
are some example spectra for the wavelength ranges used to 
determine the stellar parameters. The stars in these plots have 
a signal to noise in the middle of our observed range.

The full spectrum was used in determining 
the velocities of the program stars by employing 
a method of steepest descent to find a minimum of the sum of the 
squares of the differences between the data and synthetic spectra.  
Once a radial velocity was found, 
it was transformed into different rest frames for easier 
comparison with both previous observations and different models.  In 
particular, the values given by \citet{sch10} were adopted to 
correct to the local standard of rest (v$_{LSR}$), and a solar circular 
velocity of 225~km~s$^{-1}$ was used to determine velocities in the 
Galactic rest frame (v$_{GSR}$).

The method of extracting all the relevant information for a given 
star began with determining its radial velocity.  
After finding this initial velocity, 
an initial set of stellar parameters were determined.  In order to 
accomplish this task, a Bayesian approach was implemented.  Specifically, 
we used a Bayesian posterior distribution defined as
\begin{eqnarray*}
P_A \propto L * P_B
\end{eqnarray*}
where $P_A$ and $P_B$ are the posterior and prior probabilities 
respectively and $L$ is the likelihood which we define as
\begin{eqnarray*}
L &=& e^{-\chi^2 / 2},\\
\chi^2 &=& \sum_i \left(\frac{f_{o,i}-f_{s,i}}{\sigma_i}\right)^2
\end{eqnarray*}
where the sum is over the observed data points in the wavelength 
range of interest, 
the subscript $o$ denotes an observed value, the subscript 
$s$ denotes the synthetic value at the same wavelength and the 
flux, $f$, has been normalized.  The synthetic flux values 
are evaluated at the observed wavelength value using a cubic 
spline interpolation.  The variable $\sigma_i$ is the normalized 
noise estimate of the data determined from the non-normalized 
flux $F$ at each data point and defined as
\begin{eqnarray*}
\sigma_i = \sqrt{\frac{2}{F_{o,i} * GAIN}}
\end{eqnarray*}
which is derived from the equation given by \citet{mas92} for 
'ugly data.'

We sequentially determine the stellar parameters 
and use each new result as a prior distribution for the next.  
Since we require some initial prior probability to get this method 
started, we use the precise photometry available 
from the SDSS database along with the effective 
temperature's dependence on $(g-r)_\circ$ color given by \citet{fuk11} 
as a prior probability distribution for our effective temperature 
determination.  Specifically, we use a Gaussian centered on the 
temperature given by their equation using the $(g-r)_\circ$ color of the 
star with a width of 250~K (which is the spacing of our synthetic 
grid).  In \citet{fuk11} they quote an rms scatter of 93~K, but we 
only want this prior to guide our spectroscopic temperature 
determination so we 
use the wider width of the grid spacing so as not to force our 
result to the photometrically determined value.  
We then compute the likelihood at each synthetic 
grid point using the H$_\beta$ region of the spectrum.  Combining 
the likelihoods and prior probability distribution gives us the 
posterior probability distribution for the effective temperature.  We 
next determine the posterior probability distribution for the 
metallicity of the star using the effective temperature's 
posterior probability as the prior probability and the 
likelihoods calculated based on the wavelength ranges mentioned 
above and shown in Table \ref{tab:ranges}.  Finally, we use both 
the effective temperature and metallicity posterior probabilities 
as the prior probability for the surface gravity determination 
and calculate the likelihoods in the same manner as the other 
two parameters.  The actual parameters (and their associated errors) 
are determined from the normalized posterior probability 
distributions.  The most likely values (i.e. the parameters) are 
just the weighted averages of each distribution and the errors are the 
weighted standard deviations.

There are numerous ways that one can use the Bayesian approach 
to determine the stellar parameters in a similar manner.  We have 
chosen the approach described above because the parameter we are 
most interested in is the surface gravity.  This parameter is 
difficult to determine and the added information from 
the other two parameters in the form of prior probabilities 
improves our results.  This could result in the other two 
parameters being less certain, but as we show in the next section, 
the overall properties of the RC sample are consistent with the 
expectations of these types of stars in the Sgr tidal debris.

After finding an initial set of stellar parameters, we perform 
another round of radial velocity and parameter determinations.  
While we don't expect the values to change significantly, this 
second iteration of finding both the radial velocity and stellar 
parameters insures that we have the best possible values.  In 
particular, we want to use the closest matching grid points to 
determine the radial velocity and we want the best radial 
velocity to determine the final parameters.  Therefore, we adopt 
the second iteration values as our parameters although there 
is little change between the two sets of values.

\subsection{Calibration and Validation}\label{ssec:val}
In order to test the validity of our parameter determinations, 
we used the results of the stars in our observed sample that 
were also spectroscopically observed by the SDSS.  This is a 
rigorous test of our method and software since we are able to 
run our routine on the spectra obtained both by the SDSS and on WIYN. 
A comparison of the parameters determined with our routine using 
SDSS spectra to those determined by the SEGUE stellar parameter 
pipeline (SSPP) is a direct test of the accuracy of our parameter 
estimation method.  Furthermore, a comparison of parameters 
from the WIYN data not only our depends on our routine, but the 
different data sets as well.  There were approximately 70 stars in 
common and we found good agreement between all three sets of parameters 
with no large offsets or systematic trends.  
When running our software on SDSS spectra and comparing the results 
with those of the SSPP, we have mean absolute differences of 110~K, 
0.2~dex, 0.1~dex, and 7~km~s$^{-1}$ in effective temperature, surface 
gravity, metallicity, and radial velocity, respectively.  The standard 
deviations of the differences in the effective temperature, surface 
gravity, metallicity, and radial velocity are 75~K, 0.2~dex, 0.2~dex, 
and 3~km~s$^{-1}$, respectively.
When comparing the parameters from our WIYN spectra to those of 
the SSPP we find mean absolute differences of 
25~K, 0.05~dex, 0.2~dex, and 8~km~s$^{-1}$ and standard deviations of 
100~K, 0.4~dex, 0.7~dex, and 9~km~s$^{-1}$ for the effective temperature, 
surface gravity, metallicity, and radial velocity, respectively.

\begin{figure}[tb]
\plotone{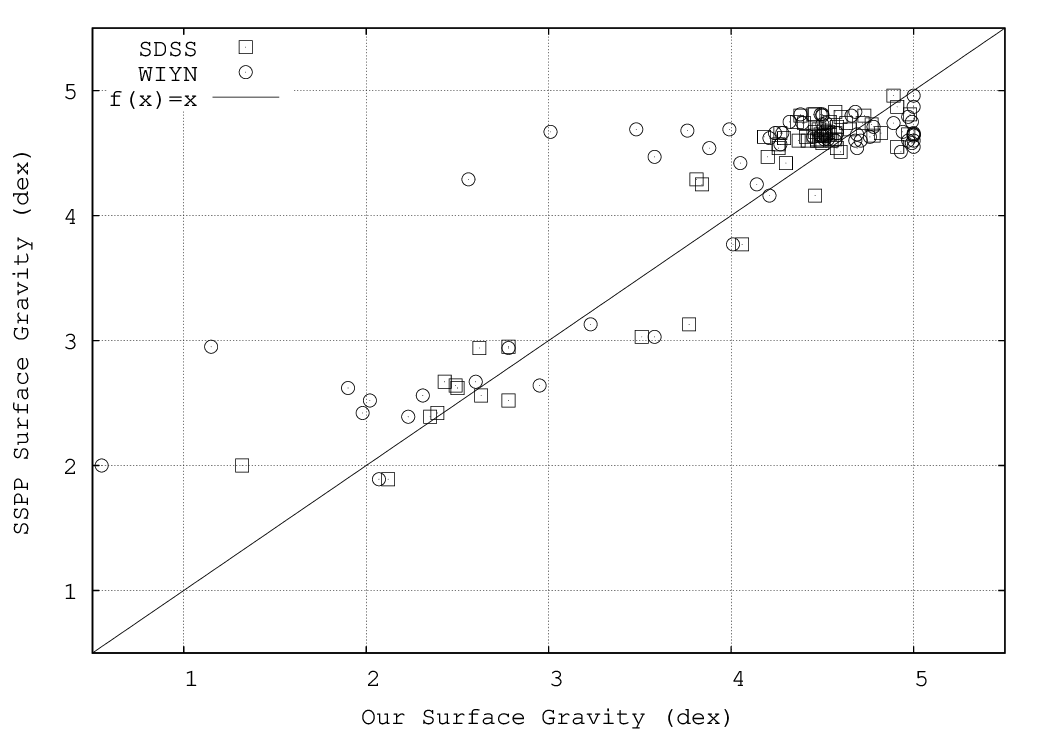}
\caption{A comparison of the surface gravity values determined 
from SDSS (squares) and WIYN (circles) spectra 
using our stellar parameter determination routine as described 
in the text.  Our values 
are along the horizontal axis and the SEGUE stellar pipeline 
values are along the vertical axis.}
\label{fig:logg-comp}
\end{figure}
\begin{figure}[tb]
\plotone{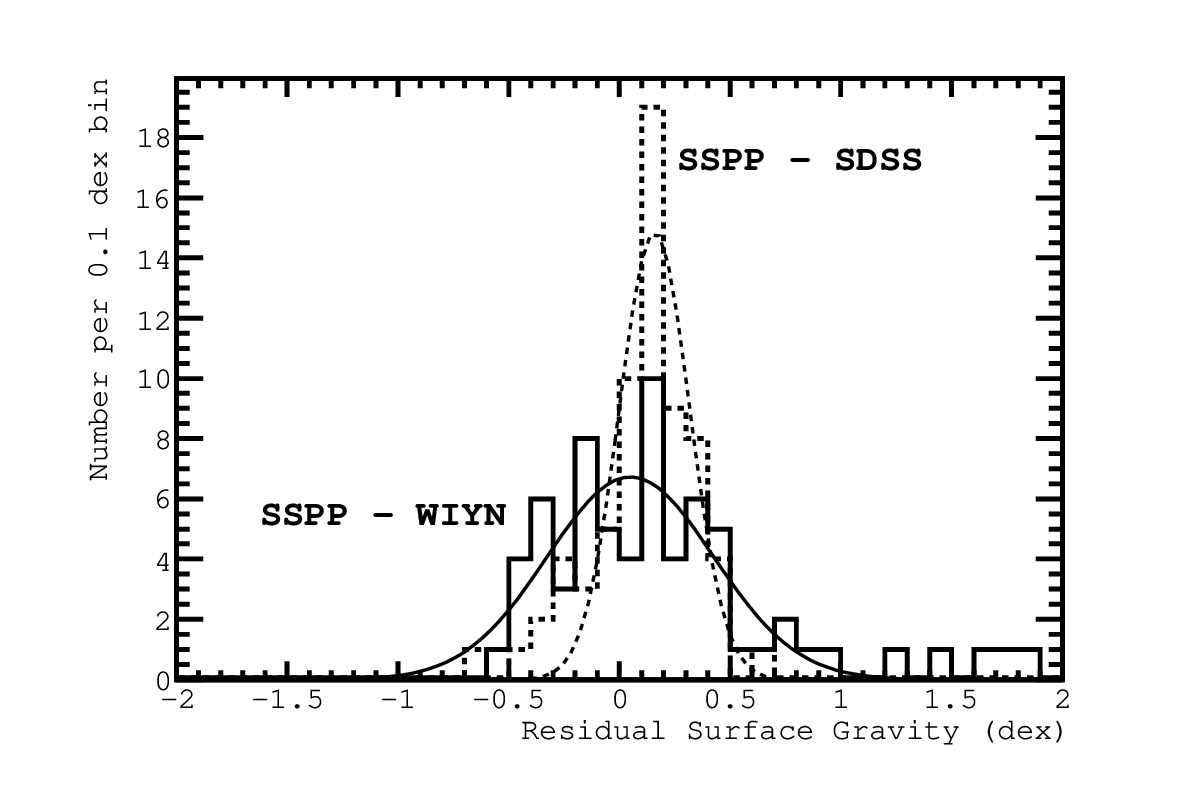}
\caption{Residual histograms of surface gravities determined from 
SDSS (dashed) and WIYN (solid) spectra using our stellar 
parameter determination routine subtracted from SEGUE stellar 
pipeline values.  The Gaussian fits that are overlaid have widths 
of 0.2 and 0.4~dex for the SDSS and WIYN data, respectively.}
\label{fig:logg-hist}
\end{figure}

Since we are most interested in the surface gravities because of the 
ability to separate dwarf stars from giant stars, a comparison 
plot and residual histograms for this parameter are given in Figures 
\ref{fig:logg-comp} and \ref{fig:logg-hist}, respectively, to better 
show the quality of our parameter determination.  The 
Gaussian fit to the WIYN data in Figure \ref{fig:logg-hist} 
finds $\sigma$~=~0.4~dex, which we take as a measure of error for 
our entire sample.  On a star by star basis the errors on the 
surface gravities are slightly higher (on average 0.5~dex) but not 
significantly so.  Our routine returns surface gravities for the 
WIYN spectra which are not quite as precise as values from the 
SEGUE stellar pipeline, but this is expected since we use spectra 
with relatively low S/N.

\begin{figure}[tb]
\plotone{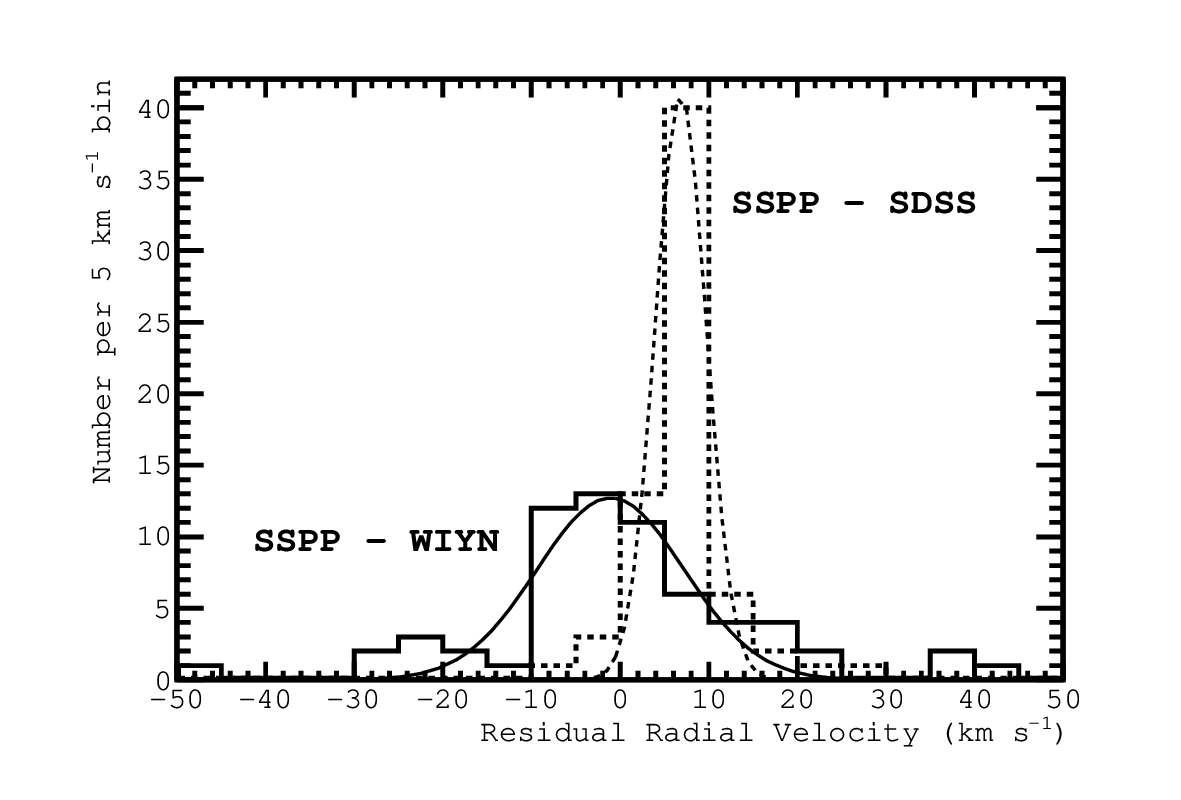}
\caption{Residual histograms of radial velocities determined from 
SDSS (dashed) and WIYN (solid) spectra using our stellar parameter 
determination routine subtracted from SEGUE stellar pipeline 
values.  The Gaussian fits that are overlaid have widths of 
3 and 8~km~s$^{-1}$ for the SDSS and WIYN data, respectively.  Note 
that the WIYN data has been calibrated by accounting for a small 
offset while the SDSS data has not.}
\label{fig:rv-hist}
\end{figure}

The radial velocities from the WIYN spectra have a 
slight offset ($\sim$8~km~s$^{-1}$) with respect to the SDSS pipeline 
values.  We use this offset to calibrate our radial velocities.
\begin{figure}[tb]
\plotone{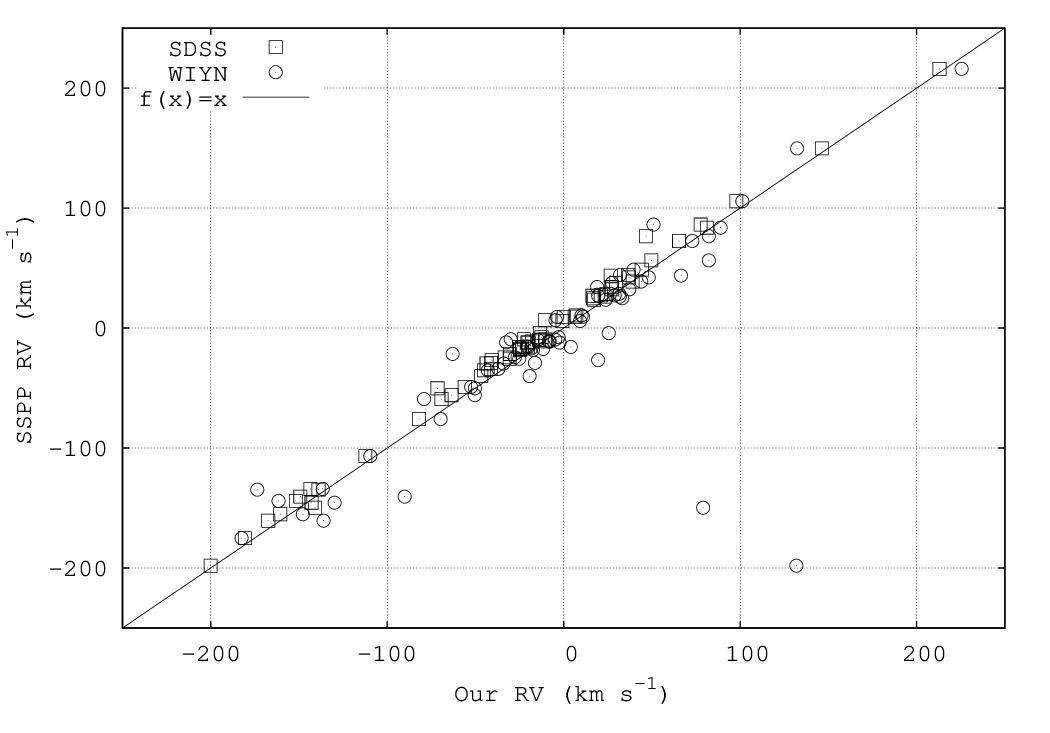}
\caption{A comparison of the radial velocities determined from 
SDSS (squares) and WIYN (circles) spectra using our stellar 
parameter determination routine as described in the text.  Our 
values are along the horizontal axis and the SEGUE stellar 
pipeline values are along the vertical axis.}
\label{fig:rv-comp}
\end{figure}
In Figures \ref{fig:rv-comp} and \ref{fig:rv-hist} one can see the 
agreement between our results and the SEGUE stellar 
pipeline.  In these plots, the values from the WIYN spectra 
have been calibrated to account for the small offset while 
the values from the SDSS spectra have not.  The published 
pipeline velocities also have a small offset applied 
(see section 2.2 from \citet{lee08} and the spectroscopic 
caveats webpage\footnote{http://www.sdss3.org/dr8/spectro/caveats.php} 
for DR8), which is similar to the difference we find between the WIYN 
stars and the stellar pipeline values.  Further, the there is no 
obvious systematic across a velocity range spanning 
roughly 500~km~s$^{-1}$ so calibrating our velocities using this small 
offset should not affect our results significantly.

As a consistency check, we look at the properties of our 
selected RC sample and how it compares to previous results.  
For our RC sample we find an average T$_{eff}$ of 5240~K with a 
standard deviation of 220~K.  The empirical calibration of 
\citet{alo99} (Equation 6 in their Table 2) for F-K giants 
gives a temperature of 5480~K with a standard deviation of 125~K 
in the middle of our $(V-I)_\circ$ color range ($(V-I)_\circ=0.9175$).  
Using the color-T$_{eff}$ relation of \citet{mcw90} (Table 6) we find 
a temperature of 5580~K with an uncertainty of 80~K using the 
middle of our color range.  Our T$_{eff}$ values agree better 
with SSPP values, but this comparison suggests that our 
temperatures may be systematically too cool by up to 300~K.  
Our average metallicity for the RC stars is -0.5~dex, which is in 
agreement with previous metallicity estimates of the Sgr dwarf 
galaxy system (see, e.g., \citet{cho07} and their discussion in 
section 2 of the paper).  The standard deviation on the mean of 
our metallicity is 0.7~dex which is primarily due to the low 
resolution and S/N of our spectra.

In summary, we have implemented a statistically sound and robust 
method for determining relevent properties of stars using a 
Bayesian approach.  This method provides results that 
are in good agreement with results from the SEGUE stellar 
pipeline for stars observed both on WIYN and by the SDSS.  We 
are therefore confident that we can use the results of this 
routine to separate populations of stars based on their 
stellar parameters and kinematics.

\section{Results}\label{sec:results}
As was described in section \ref{ssec:samp}, our selection 
criteria allowed for the inclusion of quite a significant 
background, namely the disk of our Galaxy.  Since we observed 
stars spectroscopically, however, we were able to take advantage 
of the fact that the stream and disk stars are in different 
stages of the stellar life cycle.  This coupled with the velocity 
information provides discrimination between the stream and disk 
stars.  We will first describe the procedure used to select the 
stream stars from our full sample and will follow that with 
the results these stars provide us.

\subsection{Separating the Sample}\label{ssec:sepsamp}
\begin{figure}[tb]
\plotone{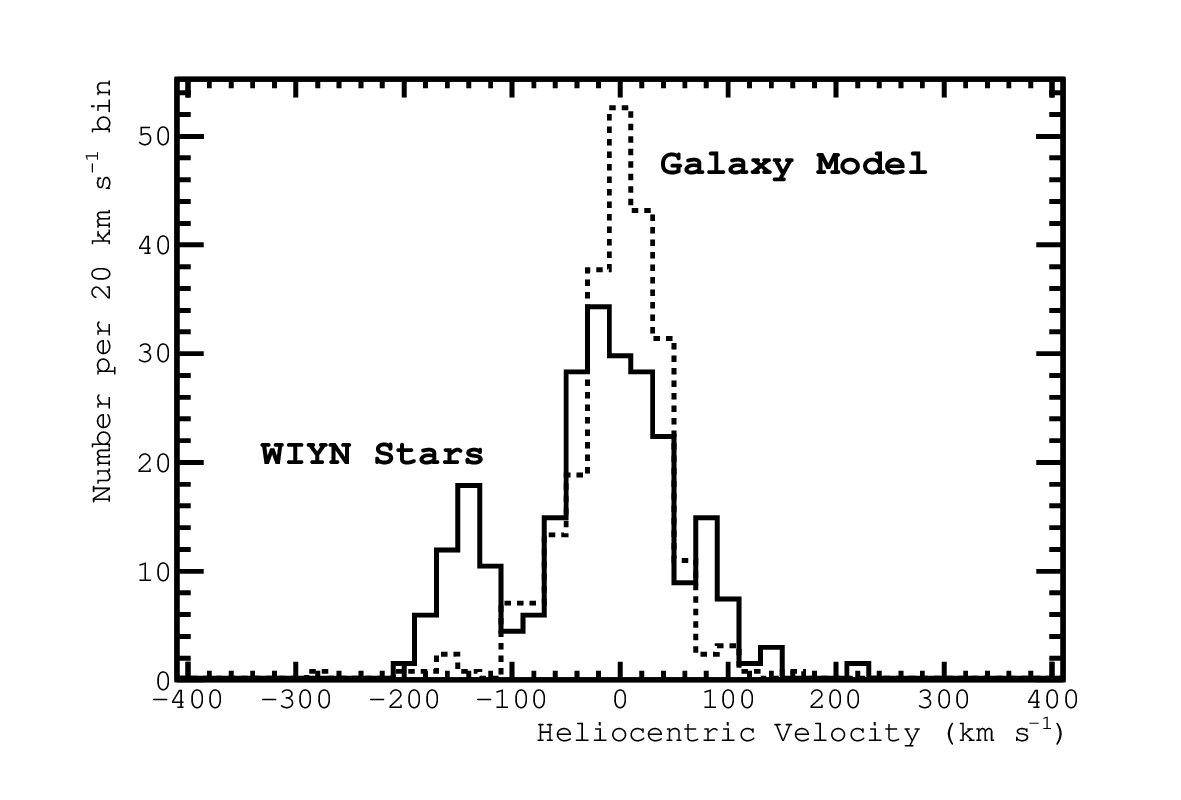}
\caption{Histogram of the heliocentric velocities of all stars 
observed in our 5 WIYN fields (solid) and those from the Besan\c{c}on 
model (dashed).  The contamination from the disk in the WIYN data 
is the large peak with a mean of $\sim$0~km~s$^{-1}$  that 
overlaps with the expected disk velocities from the model.  Stream 
stars are clearly visible as the smaller peak with velocities 
of $\sim$~-150~km~s$^{-1}$.}
\label{fig:full-vel}
\end{figure}

\begin{figure}[tb]
\plotone{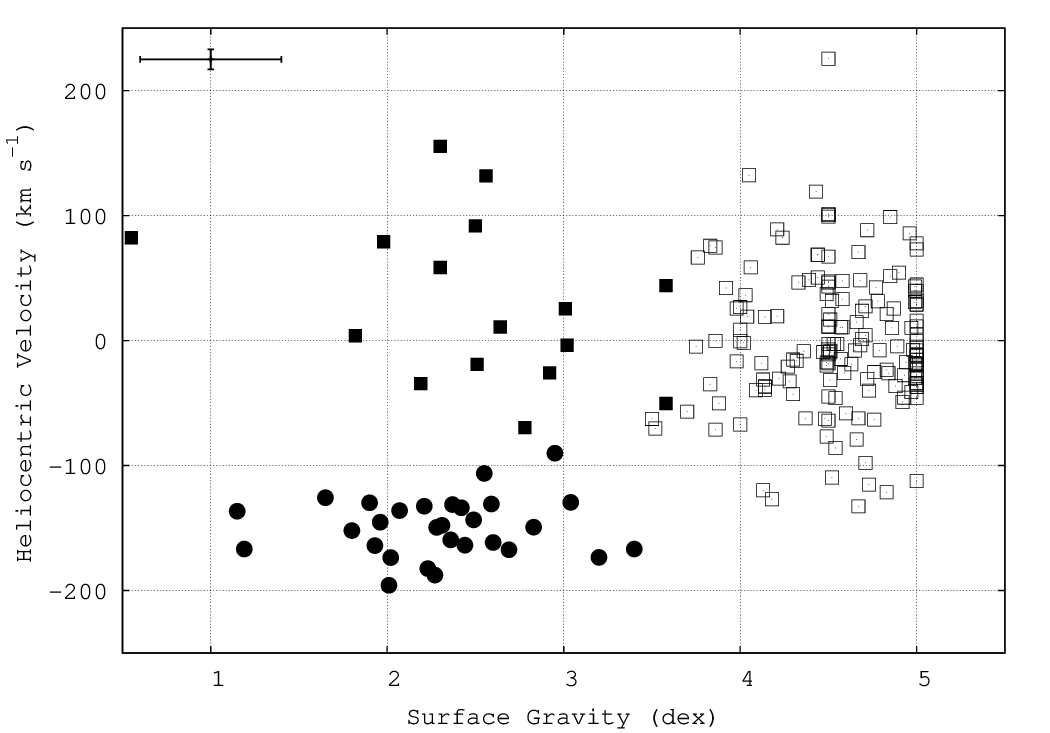}
\caption{Heliocentric velocity versus surface gravity for our 
full sample of observed stars.  The filled symbols are stars that 
satisfy the condition $\log g$~$<$~3.0~dex within errors.  Circles 
are stars which are also within 3$\sigma$ of the fitted Gaussian 
mean for the group of stars with velocities of $\sim$~-150~km~s$^{-1}$.  
Representative errors for both parameters are given in the upper 
left corner.}
\label{fig:vVlogg}
\end{figure}

Figure \ref{fig:full-vel} shows a histogram of heliocentric 
velocities for all the stars observed in our 5 WIYN fields that 
have $g_\circ$ magnitudes fainter than 16.5 as well as stars 
obtained from the Besan\c{c}on stellar population synthesis model 
of the Galaxy\footnote{\url{http://model.obs-besancon.fr}} \citep{rob03}
using the same selection criteria as the observations.
In this histogram one clearly sees disk velocities in our WIYN 
sample match those from the model ($\sim$0~km~s$^{-1}$) and are 
the dominant component.  It is important, however, to note 
that even in the full sample there is a noticeable kinematic 
signature of stream stars in the distribution.  In the plot 
of heliocentric velocities as a function of surface gravity 
for our full sample of stars shown in Figure \ref{fig:vVlogg}, 
one sees exactly what is expected if the dynamically cold 
component of the kinematic distribution is due to RC type 
stars in the Sgr streams: a large number of stars with 
low velocities ($\sim$0~km~s$^{-1}$) and high surface gravities 
($\gtrsim$~4~dex in $\log g$) as well as a population with lower 
surface gravities ($\lesssim$~3.0 dex) and very different 
velocities ($\sim$~-150~km~s$^{-1}$).
The former group is undoubtedly the disk stars in our sample and 
the latter is the RC stars in the tidal streams of the Sgr.  
In Figure \ref{fig:vVlogg} the sample has been separated with 
a cut of $\log g$~$<$~3.0~dex where stars that agree with 
this bound including errors we label as possible stream stars 
and those not satisfying it are labeled as disk dwarfs.  As 
was mentioned in section \ref{ssec:val}, the comparison 
with stars having both SDSS and WIYN spectra gives us a typical 
surface gravity error estimate of 0.4~dex and our individual 
error estimates for the WIYN stars give us a typical error 
of 0.5~dex.  One can see in Figure \ref{fig:vVlogg} that 
this is indeed the case by examining the stars with surface 
gravities between 3.0 and 3.5~dex to see which have been 
included in our low surface gravity sample.  Representative errors 
for both our surface gravities and our radial velocities 
are shown for reference in the upper left corner of the figure.

\begin{figure}[tb]
\plotone{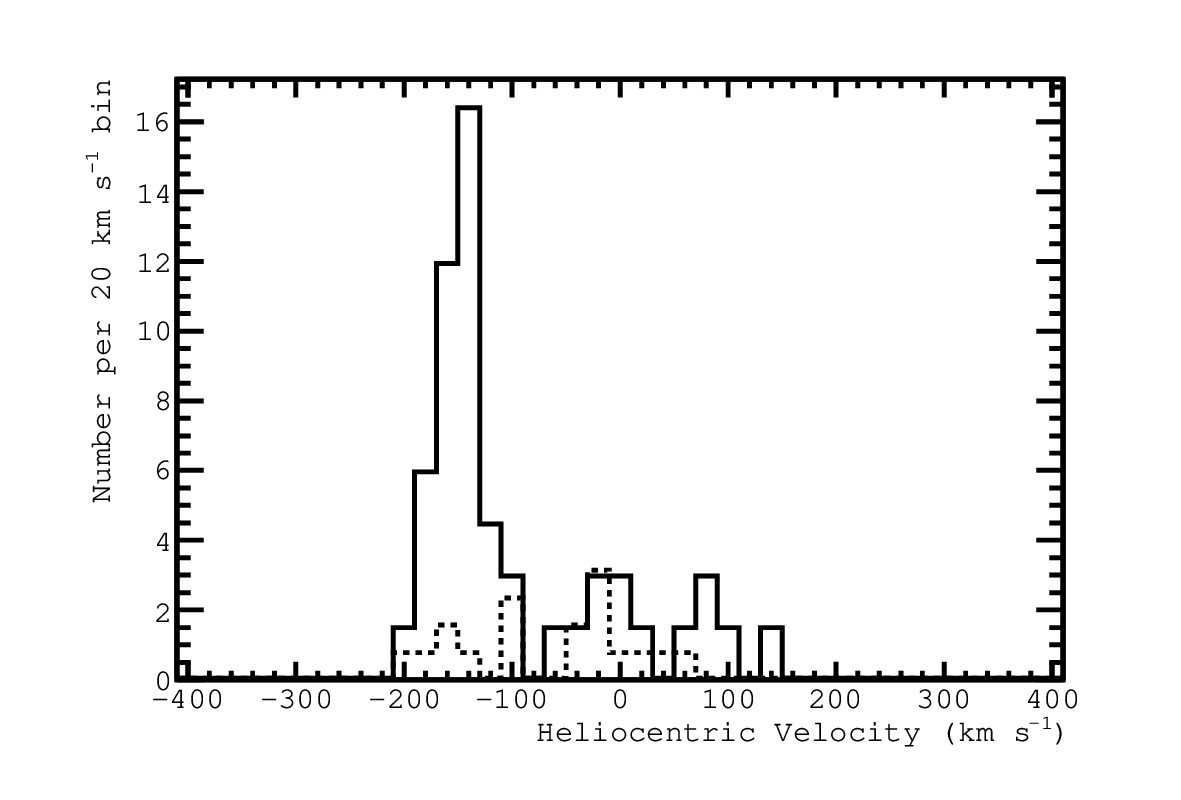}
\caption{Histograms of heliocentric velocity for our observed data 
(solid) and the Besan\c{c}on model data (dashed) which satisfy 
the surface gravity cut described in the text and have $g_\circ$ 
magnitudes fainter than 16.5.}
\label{fig:cutVels}
\end{figure}

Before determining a density, we must determine the 
quality of our sample of RC stars with regard to all 
possible contaminants.  This portion of the H-R diagram contains 
several stellar types such as older stars including the metal 
poor K giants, the metal weak RHB stars and of course, potential 
contamination from the large population of main sequence stars.  We 
have shown that by removing the higher surface gravity stars from 
the sample the contamination of disk type stars is minimal.  To 
determine the level of contamination from the halo we use the 
Besan\c{c}on model to generate a sample of Galactic stars (including 
all components) from our observed area of the sky and apply the 
surface gravity cut to eliminate disk dwarfs.  Histograms 
of heliocentric velocities for our data and the Besan\c{c}on model 
data which satisfy the surface gravity selection cut can be found 
in Figure \ref{fig:cutVels}.
The Besan\c{c}on model halo stars, chosen from our $\log g$ cut, are 
quite different from the RC sample.  The RC stars show a large peak 
with kinematics that are consistent with other publications of the 
trailing arm of the Sgr in this location (as shown below).  Other than 
the primary stream peak, our sample only shows one very small group of 
stars centered under the disk velocity distribution.  We expect that 
this small peak may 
be a very small number of contaminating disk and/or halo stars.  To 
estimate the level of contamination from misidentified dwarf stars 
we look at the number of outliers in Figure \ref{fig:logg-comp}.  
There are five stars that have SSPP surface gravities greater than 
4.0~dex and are different from our surface gravities by 0.75~dex 
(which would put them within or near our surface gravity criteria).  
Scaling this number by the fraction of observed stars in our sample 
(169 for stars fainter than $g_\circ$~=~16.5) to comparison stars 
(67) gives us approximately a dozen (12.6) 
stars, which is exactly how many stars we find outside the narrow 
peak corresponding to the Sgr stream velocity in Figure 
\ref{fig:cutVels}.

We note that there could also be a small number of contaminating K 
giants in the tidal streams (i.e. actual Sgr stream stars and 
not halo stars) as well since the Sgr has an older and more 
metal poor population as seen in the CMD of \citet{lay00}.  The 
number of metal poor giants is relatively low, however, and 
especially so when focusing on a small magnitude range as 
we do in the following.  Further, even though \citet{bel06} find 
that the relative number of BHB stars (older population) are 
enhanced in the tidal streams, they find that BHB stars still only 
account for 15\% of the total BHB and RC stars.  When comparing the 
number of K giants within our color-magnitude selection criteria to 
BHB stars in the Sgr proper (using the counts of BHB and metal 
poor giants from \citet{lay00}), the expected 
number of K giants is a factor of two less than the number 
of BHB stars.  Assuming the 15\% BHB contribution from \citet{bel06}, 
the percentage of K giants in the stream should be at most 
$\sim$~7.5\% of the RC sample.

\begin{figure}[tb]
\plotone{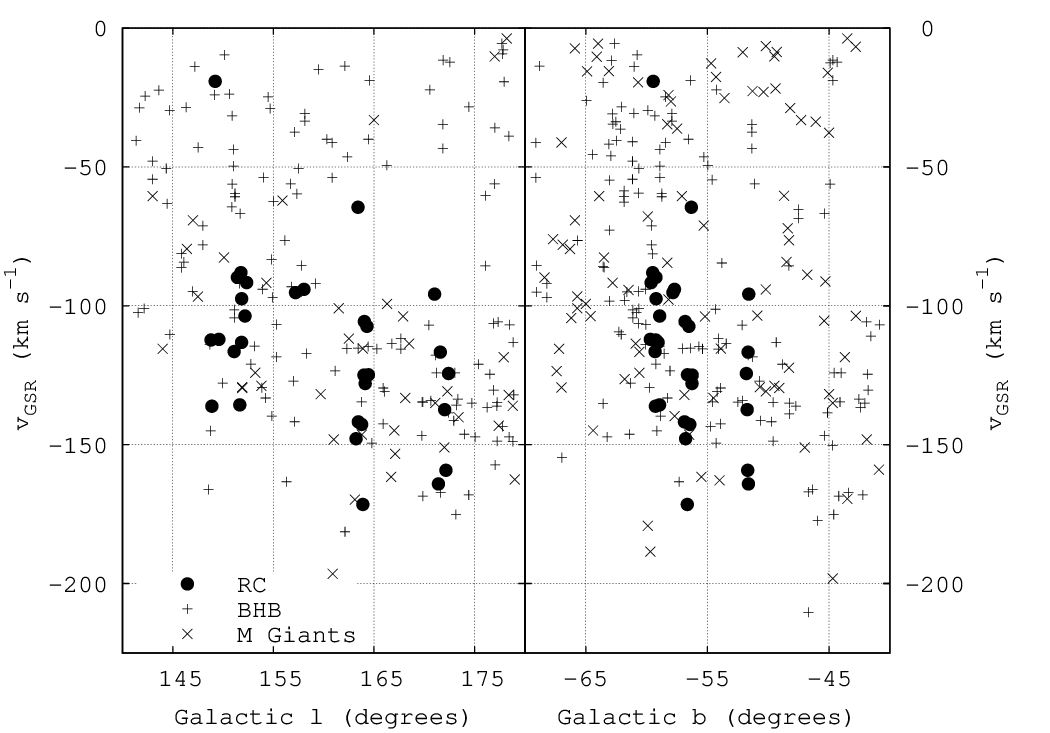}
\caption{GSR velocities of our RC sample (filled circles), the 
BHB sample (+) from \citet{yan09} and the M giant sample 
(x) from \citet{maj04} as a function of Galactic $l$ (left) and 
$b$ (right).}
\label{fig:compOther}
\end{figure}

To be sure that we are indeed seeing the trailing arm of the Sgr 
tidal streams, we compare our results 
with those of previous works.  Figure \ref{fig:compOther} 
shows that the GSR velocities for stars with low surface gravities 
match very well with those from \citet{yan09} and \citet{maj04}.

\begin{figure}[tb]
\plotone{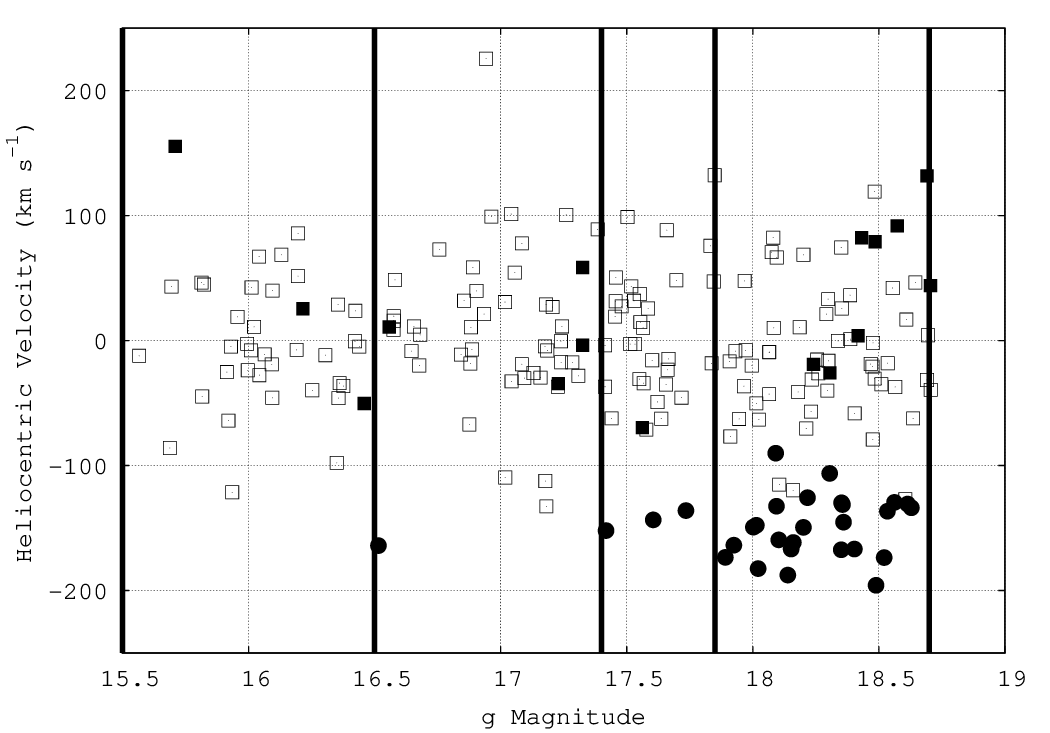}
\caption{Heliocentric velocity versus $g_\circ$ magnitude for 
our full sample of observed stars.  The filled symbols are stars 
that satisfy the condition $\log g$~$<$~3.0~dex within errors.  Circles 
are stars which are also within 3$\sigma$ of the fitted Gaussian 
mean for the group of stars with velocities of $\sim$~-150~km~s$^{-1}$.  
The vertical lines correspond to $g_\circ$ magnitudes of 16.5, 17.4, 
17.85 and 18.7.}
\label{fig:vVgmag}
\end{figure}
\begin{figure}[tb]
\plotone{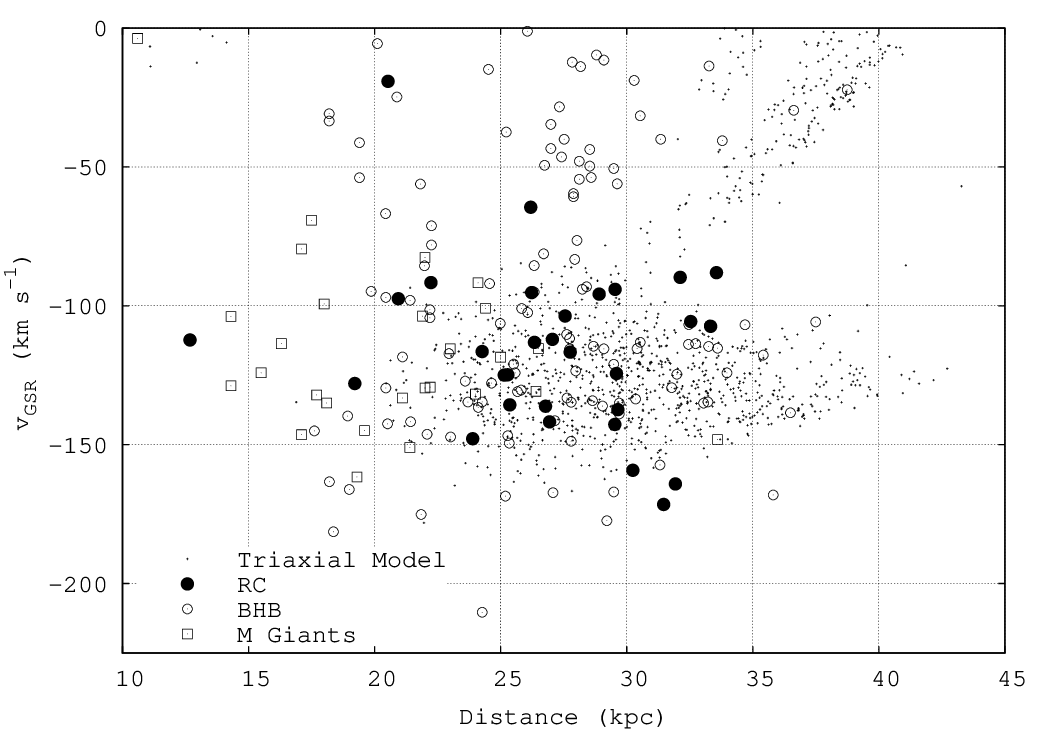}
\caption{GSR velocity versus distance for our RC stars (filled 
circles), the BHB sample (open circles) from \citet{yan09}, the 
M giant sample (open squares) from \citet{maj04} and the triaxial 
model (points) of \citet{law10}.}
\label{fig:compDist}
\end{figure}

In Figure \ref{fig:vVgmag} we show how the heliocentric 
velocities vary with $g_\circ$ magnitude.  The same surface 
gravity cut has been applied in this plot and one sees that 
not only do the low gravity stars have similar velocities, 
they also have similar magnitudes.  This point is important 
because the RC is a part of the HB and we expect that their 
absolute magnitudes should all be roughly the same.  Therefore, 
using $g_\circ$ as a proxy for the distance, our low surface 
gravity sample has a large group 
of stars with similar distances according to Figure 
\ref{fig:vVgmag}.  Also in the plot one notices several 
vertical lines which correspond to possible magnitude 
limits which will be used in the density calculations.  These 
limits are at $g_\circ$ magnitudes of 16.5, 17.4, 17.85 and 18.7 
(which is the magnitude limit of our observations).  In 
Figure \ref{fig:compDist} we see that our RC 
distances are similar to the previous results 
of \citet{yan09} and \citet{maj04} by adopting an absolute 
magnitude of $g_{abs}$~=~1.0.

We find no evidence of the leading arm in this area of the streams 
and at these distances.  The leading arm would have velocities 
of $\sim$~+150~km~s$^{-1}$ according to the models and we see no 
obvious kinematic feature other than that matching the trailing arm.  
The leading arm in our observed area would be much farther along the 
orbit from the Sgr compared to the trailing arm, which means the stars 
would have been pulled from the galaxy much earlier in the 
interaction history with our Galaxy.  This suggests, qualitatively, 
that the RC stars were not loosely bound to the Sgr during that time.

Since our color selection isolates the region of the RC 
seen in the Sgr proper and we see stars with low gravities 
that have similar velocities (which are different from 
any Galactic component in the same direction) and similar 
magnitudes, we are confident that these stars are RC stars 
in the trailing arm of the tidal streams of the Sgr.

\subsection{Density of RC Stream Stars}\label{ssec:density}
The number of stars observed with different $g_\circ$ magnitude 
limits can be found in Table \ref{tab:counts}.
\begin{table*}[tb]
\begin{center}
\caption{Star Counts\label{tab:counts}}
\begin{tabular}{ cccccc }
\tableline
{\bf Bright} & {\bf Observed} &
  \multicolumn{2}{c}{{\bf RC Stars}} &
  \multicolumn{2}{c}{{\bf Model Stars}} \\
{\bf Limit} & {\bf Fraction} & Scaled & Integral & Scaled & Integral \\
\tableline
$g_\circ >$ 16.5  & 169/252 & 41.8 & 41.5 & 5.5 & 4.9 \\
$g_\circ >$ 17.4  & 122/169 & 37.4 & 37.0 & 3.9 & 3.7 \\
$g_\circ >$ 17.85 &  87/114 & 31.4 & 32.3 & 0.8 & 0.8$^*$ \\
\tableline
\end{tabular}
\end{center}
\end{table*}
As can be seen in the table, our observations probed up to 76\% of 
the stars with clean SDSS DR6 photometry in this region of the sky 
with these color and magnitude cuts.  The candidate 
selection and placement of stars on fibers of the instrument 
is more or less completely random\footnote{Priority can be put 
on certain stars but that does not guarantee their placement.  
It only sets an order the software uses to attempt to configure 
the field and makes little practical difference.} 
so we assume our sample has no selection bias.  A small amount 
of metallicity bias is possible since the color of the RC is 
affected by metal abundance.  However, the width of the color selection 
window is such that we expect this bias to be inconsequential.  This 
allows us to scale our observed results up to what would be expected 
if we had observed every possible star in this area of the sky.

\begin{figure}[tb]
\plotone{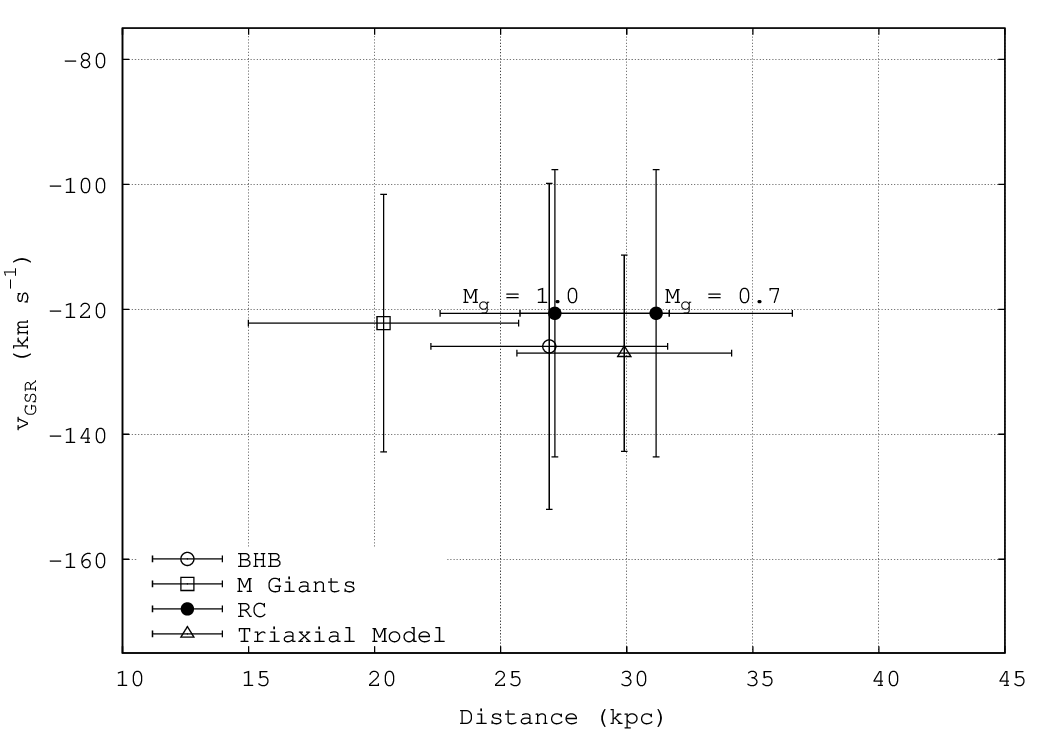}
\caption{Mean (point) and standard deviation (error bar) of the GSR 
velocities and distances for the BHB sample (open circle) from 
\citet{yan09}, the M giant sample (open square) from \citet{maj04}, 
the triaxial model (open triangle) of \citet{law10} and our RC stars 
(filled circles) using two absolute magnitudes.  All samples are 
constrained to our observed area and velocities of 
-225~$<$~v$_{GSR}$~$<$~-75.}
\label{fig:aveDist}
\end{figure}

With this information we are now ready to compute the density of RC stars 
in the trailing arm of the Sgr.  
Each circular WIYN field has a diameter of 1$^\circ$ and we must 
multiply by the 5 fields to get a total area of our observations. 
The depth of the observation was established by using an estimate of the 
appropriate absolute $g$ magnitude.  In \citet{yan09} an absolute magnitude 
of $g_{abs}$~=~0.7 was used for BHB stars which could also be 
adopted here.  However, the RC has a higher abundance than the BHB and 
this will affect the absolute magnitude of these HB stars.  We determine 
the absolute magnitude for the Sgr RC using the measurements of 
\citet{lay00}.  In their work they give the location of the RC in their 
CMD and measure a distance modulus to the Sgr using RR Lyrae stars.  Using 
this information and their reddening and extinction values then converting 
to SDSS magnitudes we find M$_g$~=~1.0~$\pm$~0.1 for the absolute magnitude 
of the RC.  One could also use the age and metallicity for the 
intermediate age stars given by \citet{lay00} to find the absolute 
magnitude based on the local RC absolute magnitude and a correction from 
globular clusters (such as in \citet{per03}).  The same magnitude is found 
using this method as the above within errors.  We calculate the density 
using both the absolute magnitude derived from the 
measurements of \citet{lay00} and the one adopted by \citet{yan09} for 
the BHB in the following, but note that our derived value more closely 
matches the properties of the RC.  In Figure \ref{fig:aveDist} we plot 
the mean and standard deviation of the GSR velocities and distances 
for our RC sample, the \citet{yan09} BHB sample, the \citet{maj04} M giant 
sample and the triaxial halo model of \citet{law10} in our observed area.
One sees that using an absolute magnitude of M$_g$~=~1.0 we get very good 
agreement with both the BHB and model stars.  The M giants look to be 
systematically closer but this is likely due to the limiting magnitude 
of their sample.  The volume encompassing 
our sample was computed by determining the volume of a cone 
truncated at the bright end of our sample.

We determine the total number of stream stars using two methods.  
The first method we use is simply counting the stars which satisfy 
our surface gravity criterion ($\log g$~$<$~3.0~dex within errors) and 
have heliocentric velocities that agree with the models and previous 
measurements (-200~km~s$^{-1}$~$<$~v~$<$-100~km~s$^{-1}$).  The other 
method we use is to fit the kinematic peak of the stars satisfying 
the surface gravity criterion with a Gaussian function 
and find the integral of this function out to 3$\sigma$ from 
the mean.  The fitted mean and $\sigma$ values for our sample of 
RC stream stars for different magnitude limits can be found 
in Table \ref{tab:veldisp}.
\begin{table}[tb]
\begin{center}
\caption{Fitted mean and dispersion of heliocentric velocity 
of RC Sgr stream stars.\label{tab:veldisp}}
\begin{tabular}{ ccc }
\tableline
{\bf Bright} & {\bf $\mu$} & {\bf $\sigma$} \\
{\bf Limit}  & (km~s$^{-1}$) & (km~s$^{-1}$) \\
\tableline
$g_\circ >$ 16.5  & -148 & 21 \\
$g_\circ >$ 17.4  & -147 & 21 \\
$g_\circ >$ 17.85 & -148 & 24 \\ 
\tableline
\end{tabular}
\end{center}
\end{table}
Both methods should (and do) give similar results, but the latter is 
not affected by the arbitrary choice of defining velocities for 
the stream stars, rather it is determined directly.  Contamination 
from halo K giants is determined in a similar manner (both with 
fits and from raw counts) from the Besan\c{c}on model data.  The 
RC counts are then scaled by our observed fraction and the model 
stars are scaled to our actual observed area to get the total 
number of RC and model (halo) stars in this area of the 
sky.  To find the actual number of expected RC stars in our 
sample we subtract the number of model (halo) stars from our 
counts of RC candidates.  We use Poisson noise as an error estimate 
for our star counts on both the model (halo) and RC candidate 
samples.  A table showing the star counts with different limiting 
magnitudes can be found in Table \ref{tab:counts}.

It is evident from the table that both the number of RC stars and 
the number of model halo stars decreases as the magnitude 
limit is decreased.  However, the halo stars decrease to 
approximately a single star in the smallest magnitude range.  
In fact, there were so few stars in the model sample that a 
Gaussian fit could not be reliably determined, and therefore 
we use the raw count for the estimate of the fitted value in the 
faintest magnitude range.  
Furthermore, in Figure \ref{fig:vVgmag} it is obvious that most of 
our candidate RC stars are in the faintest magnitude range and 
those with similar velocities to the stream at brighter 
magnitudes are consistent with the number of halo giant stars 
expected in the sample.  These two facts leads us to only 
consider stars with magnitudes of 17.85~$<$~$g_\circ$~$<$~18.7 for 
our best density estimate.  This density of 
$\rho$~=~2.7~$\pm$~0.5~RC~stars~kpc$^{-3}$ as 
well as the densities of the other magnitude ranges are 
summarized in Table \ref{tab:dens}.
\begin{table*}[tb]
\begin{center}
\caption{RC Densities\label{tab:dens}}
\begin{tabular}{ ccccc }
\tableline
\multirow{2}{*}{{\bf Bright Limit}} &
 \multirow{2}{*}{{\bf Abs. Magnitude}} & {\bf Volume} &
 \multicolumn{2}{c}{{\bf Densities (stars kpc$^{-3}$)}} \\
 & & (kpc$^{3}$) & Scaled & Integral \\
\tableline
\multirow{2}{*}{$g_\circ >$ 16.5}
 & $g_{abs}$ = 0.7 & 24.0 & 1.52 $\pm$ 0.29 & 1.53 $\pm$ 0.28 \\
 & $g_{abs}$ = 1.0 & 15.8 & 2.29 $\pm$ 0.43 & 2.31 $\pm$ 0.43 \\
\tableline
\multirow{2}{*}{$g_\circ >$ 17.4}
 & $g_{abs}$ = 0.7 & 21.0 & 1.60 $\pm$ 0.31 & 1.59 $\pm$ 0.30 \\
 & $g_{abs}$ = 1.0 & 13.9 & 2.42 $\pm$ 0.46 & 2.40 $\pm$ 0.46 \\
\tableline
\multirow{2}{*}{$g_\circ >$ 17.85}
 & $g_{abs}$ = 0.7 & 17.4 & 1.76 $\pm$ 0.33 & 1.81 $\pm$ 0.33 \\
 & $g_{abs}$ = 1.0 & 11.5 & 2.66 $\pm$ 0.49 & 2.74 $\pm$ 0.50 \\
\tableline
\end{tabular}
\end{center}
\end{table*}
The final densities for each magnitude range are determined by taking 
the RC star counts (which have been scaled by our observed fraction and 
have the scaled number of model stars subtracted out) and dividing 
by the volume for the given absolute magnitude value.  
Errors on the counts are described above and are used to determine 
the overall error on the density.  The density measurements for all 
three magnitude ranges agree with each other within errors for both 
absolute magnitude values individually.

\section{Discussion}\label{sec:discussion}
We have shown that we can successfully separate RC type stars 
from a background of disk type stars using a combination of 
kinematics and stellar surface gravities obtained from low resolution 
and signal to noise spectra.  This procedure applied to an area of the 
sky in the direction of the trailing arm of the 
Sgr provides results that agree with previous measurements published on 
the stream.  In particular, the distances and kinematics of our sample 
match other populations of stars both qualitatively and 
quantitatively.  We also make the first measurement of the 
density of RC stars at a point along the tidal streams and find 
a value of $\rho$~=~2.7~$\pm$~0.5~RC~stars~kpc$^{-3}$.
This density was found using the faintest magnitude limit, star 
counts determined from the integral of the Gaussian fits to the 
kinematic peaks and using an absolute magnitude of M$_g$~=~1.0.  We 
consider these criteria the best for several reasons.  First, our 
RC candidate stars seem to be concentrated at the faintest 
magnitudes.  Second, the fit to the kinematic peak eliminates 
problems with choosing an arbitrary velocity range for the stream. 
Finally, the derived absolute magnitude matches the Sgr RC properties 
and is in good agreement with previously published distances to the 
stream.

Since the RC population of stars in the Sgr system seems to be a 
sizeable sample, one should be able to probe differences in 
densities of stars stripped from the galaxy proper.  We note that 
while BHB stars are much easier to isolate photometrically, the 
RC stars we have observed are much more dynamically cold according 
to Figure \ref{fig:compOther}.  Identifying RC stream members 
should therefore be easier because of their large number and 
cold kinematics when compared to the BHB stream stars, which are 
harder to distinguish from the general halo field.

In a paper in preparation, we will present the density measurement 
of RC stars in another area of the streams and make a comparison 
with the results found here.  This comparison coupled with previous 
(and future) investigations will give an idea of how and when 
the young population stars were stripped from the Sgr proper during 
previous orbits.  Gradients along the stream 
of PopI stars has been suggested in \citet{bel06} but more measurements 
of several different populations is needed for a complete accounting 
of the interaction history.  Detailed measurements of the distribution 
of stars along the tidal streams will also make it possible to compute 
the total mass in the stream and will provide information on the original 
distribution of these stars in the Sgr.  This is crucial for 
determining the IMF of the original galaxy.  Furthermore, determining 
the densities and locations of different populations of stars in the 
tidal streams will provide constraints on the internal dynamics of 
the original Sgr.  This information is needed for better and more 
accurate models of the Sgr interacting with our Galaxy.

The Sgr orbiting in and around the potential of our Galaxy offers a 
unique opportunity to study an interacting galactic system.  The RC 
population can be used as an effective probe to help constrain both 
the shape of the dark matter halo and the initial distribution of 
stars in the Sgr galaxy.  We have shown that this population 
of stars can be extracted from the significant background and yield 
results which are consistent with other populations of stars.  Several 
different populations of stars with different ages are needed to 
fully characterize the interaction of the Sgr with our Galaxy and the 
RC is a justifiable and observable group to use as a young/intermediate 
component.

\acknowledgments

We would like to acknowledge the NOAO for granting time and 
providing funding to make the necessary observations for 
this research.  Funding was also provided by the Texas 
Space Grant Consortium and the Sigma Xi Grants In Aid of Research 
Program.  This work was also supported by the National Natural Science 
Foundation of China under grants 11150110135, 11073026, and 11178013 
and the Chinese Academy of Sciences under grant KJCX2-YW-T22 and the 
fellowship for young international scientists.

Funding for the Sloan Digital Sky Survey (SDSS) and SDSS-II has 
been provided by the Alfred P. Sloan Foundation, the Participating 
Institutions, the National Science Foundation, the U.S. Department 
of Energy, the National Aeronautics and Space Administration, the 
Japanese Monbukagakusho, and the Max Planck Society, and the Higher 
Education Funding Council for England. The SDSS Web site is 
http://www.sdss.org/.

The SDSS is managed by the Astrophysical Research Consortium (ARC) 
for the Participating Institutions. The Participating Institutions 
are the American Museum of Natural History, Astrophysical Institute 
Potsdam, University of Basel, University of Cambridge, Case Western 
Reserve University, The University of Chicago, Drexel University, 
Fermilab, the Institute for Advanced Study, the Japan Participation 
Group, The Johns Hopkins University, the Joint Institute for Nuclear 
Astrophysics, the Kavli Institute for Particle Astrophysics and 
Cosmology, the Korean Scientist Group, the Chinese Academy of 
Sciences (LAMOST), Los Alamos National Laboratory, the 
Max-Planck-Institute for Astronomy (MPIA), the Max-Planck-Institute 
for Astrophysics (MPA), New Mexico State University, Ohio State 
University, University of Pittsburgh, University of Portsmouth, 
Princeton University, the United States Naval Observatory, and the 
University of Washington.

{\it Facilities:} \facility{WIYN (Hydra)}.

\end{document}